\documentclass[manuscript]{aastex}

\usepackage{hyperref}
\usepackage{natbib}
\usepackage{amsmath}
\usepackage{amssymb}
\usepackage{graphicx}
\usepackage{color}
\usepackage{verbatim}

\shorttitle{The GRB 030329 Radio Afterglow}
\shortauthors{Mesler et al.}

\begin{document}
\title{VLBI and Archival VLA and WSRT Observations of the GRB 030329 Radio Afterglow}


\author{Robert A. Mesler}
\affil{Department of Physics and Astronomy, University of New Mexico MSC07 4220, Albuquerque, NM 87131}
\email{meslerra@unm.edu}

\author{Ylva M. Pihlstr\"om\altaffilmark{1} \and Greg B. Taylor\altaffilmark{1}}
\affil{Department of Physics and Astronomy, University of New Mexico, MSC07 4220, Albuquerque, NM 87131}
\altaffiltext{1}{Ylva Pihlstr\"om and Greg Taylor are also Adjunct Astronomers at the National Radio Astronomy Observatory} 

\and

\author{Johnathan Granot}
\affil{Centre for Astrophysics Research, Science and Technology Research Institute, University of Hertfordshire, College Lane, Hatfield AL10 9AB, UK}



\begin{abstract}
We present VLBI and archival Karl G. Jansky Very Large Array (VLA) and Westerbork Synthesis Radio Telescope (WSRT) observations of the radio afterglow from the gamma-ray burst (GRB) of 2003 March 29 (GRB 030329) taken between 672 and 2032 days after the burst.  The EVLA and WSRT data suggest a simple power law decay in the flux at 5 GHz, with no clear signature of any rebrightening from the counter jet. We report an unresolved source at day 2032 of size $1.18\pm0.13$ mas, which we use in conjunction with the expansion rate of the burst to argue for the presence of a uniform, ISM-like circumburst medium.  A limit of $< 0.067$ mas yr$^{-1}$ is placed on the proper motion, supporting the standard afterglow model for gamma-ray bursts.
\end{abstract}




\section{Introduction}

Gamma-ray bursts (GRBs) are sudden, short-lived ($\lesssim 10^2$ s) releases of energy of on the order $10^{51}$ erg \citep{frailEA01, bloomEA03} in a region of $\lesssim 100$ km.  The central engine, thought to be a black hole with an accretion disk formed either by the collapse of a massive evolved star (collapsar) for long GRBs ($\gtrsim 2$s) or by the merger of two stellar remnants (black holes or neutron stars) for short GRBs ($\lesssim 2$s).  The large, sudden energy release produces an $e^+e^-$,$\gamma$ fireball in the form of collimated jets \citep{paczynski93}.  Variability in the rate and velocity at which material is ejected from the central engine leads to internal shocks within the jet when fast-moving shells catch up to slower ones.  The resulting $\gamma$-ray emission is responsible for the so-called prompt emission \citep{reesMeszaros94}.

Whereas prompt emission lasts for $\lesssim$ a few minutes, GRB afterglow emission can last far longer, up to years in the radio part of the spectrum, e.g. \citep{frailEA00, bergerEA03c, taylorEA04, pihlstromEA07}.  The long-lived afterglow emission is primarily synchrotron radiation is usually attributed to an interaction with the shocked external medium that exists behind the external forward shock \citep{meszarosRees97}. More recently, perplexing features observed by Swift in the early X-ray afterglow have lent weight to a possible alternative scenario wherein a long-lived reverse shock decelerates slow ejecta at the back of the original outflow as it gradually catches up with the shocked external medium \citep{genetEA07, uhmBeloborodov07}. Whether the afterglow arises from the forward external shock or from a long lived reverse external shock, an external shock origin is strongly supported by measurements of the radio afterglow image size and its temporal evolution for the late time radio afterglow of GRB 030329 \citep{taylorEA04, orenEA04, granotEA05, pihlstromEA07}.

To explain the intensity and duration of observed GRB afterglow emission, the presence of a circumburst medium must be taken into account.  External shocks naturally arise when jet material interacts with the circumburst medium, converting kinetic energy of the fireball into particle energy and luminosity \citep{meszarosRees93}.  These interactions are expected to be collisionless, mediated instead by the tangled and compressed magnetic fields that exist at the shock boundary.  Electrons accelerating along magnetic field lines at the shock boundary produce the power law emission spectrum characteristic of gamma-ray bursts \citep{meszarosRees93, katz94}.  The spectral and temporal evolution \citep{sariEA98, granotsari02} of GRB afterglows, then, are governed by such factors as the structure and dynamical evolution of the relativistic jet (for a review, see \citealt{granot07}), but also by the environment (e.g., constant density versus a windlike density profile; \citealt{chevalierLi00}).  A detailed study of the size evolution of a GRB afterglow would make it possible to better constrain the density profile of the circumburst medium.

Progenitors of long gamma ray bursts are expected to be associated with the collapse of evolved stars (collapsars) \citep{woosley93, paczynski98, FryerEA99}.  At least some GRBs should therefore exist in gas-rich environments with isotropic windlike $\rho(r) \propto r^{-2}$ density profiles out to a few tenths of a parsec, where there is a region of roughly uniform density corresponding to the stellar wind reverse shock \citep{ramirez-RuizEA05}.  Because the radio afterglow is detectable for months or years after the initial explosion, the jet would be expected to have sufficient time to traverse the stellar wind and enter the uniform shell of material that lies beyond.  Several long GRBs, however, have shown light curves that are more characteristic of a jet propagating through a uniform $\rho(r) = \rho_0$ medium for the entire observed duration of the radio afterglow.  More direct and less model-dependent methods of determining the radial density profile could be used to better constrain the progenitor as well as to test the validity of the standard fireball model.

In the simplest emission models, the flux density evolution of the GRB afterglow is related to the density of the medium in which the GRB is present \citep{waxman97}.  The afterglow from GRB 030329 was monitored by \citet{taylorEA04} and \citet{pihlstromEA07} to obtain the first well-determined expansion rate for a GRB up to 247 and 806 days after the burst, respectively.  The apparent diameter of the burst afterglow was shown to increase from $0.065 \pm 0.022$  mas to $0.172 \pm 0.043$ mas between day 25 and 83.  Thereafter, the expansion rate decreased, with the burst reaching a size of $0.347 \pm 0.090$ mas at $t = 806$ days.  The evolution of the mean apparent expansion speed of the afterglow image \citep{pihlstromEA07} suggests a transition to non-relativistic expansion after about one year (see Fig.\ \ref{beta}). Models of the afterglow expansion predicted similar sizes at day 217 and day 806 in both the uniform density and the wind density profile case (see Fig.\ \ref{models}), making it necessary to observe the afterglow at later times so that the preferred model for the GRB 030329 circumburst environment could be unambiguously determined using this method.  Here we present observations of the GRB 030329 afterglow taken up to 5.5 years after the initial burst in order to shed light on the validity of the different afterglow models.  We also seek to place constraints on the density profile of the circumburst environment, the jet dynamics and structure, and the proper motion. 

\section{GRB 030329}
The gamma-ray burst GRB 030329 was first detected by the High Energy Transient Explorer 2 (\emph{HETE-2}) satellite at 11:37 UTC on March 29th, 2003 \citep{vanderspekEA03}, and was localized in the optical bands by \citet{petersonPrice03}.  Its redshift of $z = 0.1685$ \citep{greinerEA03}, corresponding to an angular distance of 587 Mpc, makes it one of the nearest GRBs to Earth.\footnote{The redshift of GRB 030329 is found to be $z = 0.1685$.  Assuming a $\Lambda_{CDM}$ cosmology with $H_0 = 71$ km s$^{-1}$ Mpc$^{-1}$, $\Omega_M = 0.27$, and $\Omega_{\Lambda} = 0.73$, GRB 030329 is located at an angular distance of $d_A = 587$ Mpc, with 1.0 mas corresponding to 2.85 pc.}  The GRB 030329 radio afterglow is the brightest ever to have been observed, reaching a maximum flux density of 55 mJy one week after the burst at 43 GHz \citep{bergerEA03a}.  The relative proximity of GRB 030329 to the Earth and its high afterglow luminosity present an unprecedented opportunity to study the flux evolution of a GRB afterglow, as well as the rate of expansion of the burst, which was fully resolved by the VLBA at 4.86 GHz \citep{taylorEA04,taylorEA05,pihlstromEA07}.


\section{Observations and Data Reduction}


\subsection{Global VLBI}

On 21 October 2008, 2032 days after the burst, we used 9 antennas from the global VLBI Network (Effelsberg, Jodrell Bank Lowell, Medicina, Noto, Onsala, Torun, Westerbork, Green Bank, and Arecibo) to observe the 5~GHz continuum afterglow emission of GRB~030329. This array provided a baseline range between 266 and 6911~km, corresponding to fringe spacings between 1.8 and 47 mas.  The total bandwidth observed was 128~MHz (8 IFs of 16~MHz each), dual polarization. The nearby source J1048+2115 was used as the phase-reference calibrator with a 3 minute cycle (2 minutes on-source, 1 minute on the calibrator). The source was tracked for 7.5~hours in total, which included 3~hours with the Arecibo telescope.

Data were recorded at 1~Gb/s rate, except for at the GBT telescope where the data recording rate was limited to 512~Mb/s. The data were correlated at the Joint Institute for VLBI in Europe (JIVE) with a 2 seconds averaging time and 16 channels, and the GBT data was correlated as 2-bit data with the magnitude bit set to high. This correlation scheme was successfully applied to our previous high sensitivity observations (Pihlstr\"om et al.\ 2007).

Data reduction was done using the AIPS package. First the data were corrected for a slight position offset for the GBT antenna, and then for a position offset of the phase calibrator determined by the position of the ICRF source J1051+2119 observed every 30 minutes. This ensures the position is measured relative to the same points as has been done in previous VLBI observations of this source \citep{pihlstromEA07, taylorEA05, taylorEA04}. A continuum map was produced using the inner 75\% of channels in each IF, and resulted in a 6 $\mu$Jy beam$^{-1}$ RMS noise level.

A circular Gaussian was fitted to the measured visibilities and we derived an angular diameter of $1.18\pm0.13$ mas.  The error in this value comes from uncertainties in the size of the gaussian fit, and arises due to the RMS noise contained in the image.  It should not be confused with the size of the source, which is not resolved.  Rather, it is a measurement of the angular resolution of the image, and can only serve as an upper limit to the source size.  The position obtained for GRB030329 at day 2032 is R.A. = 10$^h$44$^m$49$^s$.95944 and Dec = 21$^{\circ}$31'17.4372".


\subsection{VLA and WSRT Archival Observations}
Here we present late-period 5 GHz observations, taken from the NRAO and WSRT data archives, that were made between 672 and 1773 days after the burst (see Table \ref{VLAobservations}).  The VLA data were reduced in the AIPS software package.  The bright calibrator 3C286 was used for absolute flux calibration, and J1051+213, which was separated from the source by $1.6^{\circ}$, was used to determine the complex antenna gains.  
The WSRT data were originally published in \citet{vanderHorstEA05} and \citet{vanderHorstEA08}.  WSRT data appearing in those two works covering the same time span as the VLA data that we have taken from the NRAO archive were re-analysed to ensure uniformity of analysis.  The re-analysed data were reduced in AIPS, with 3C286 serving as the primary calibrator and either 3C48 or 3C147 serving as the secondary calibrator.  Flux densities derived from the WSRT data that were re-analysed in this work appear in Table \ref{VLAobservations}.

\begin{table}[t]
\begin{center}
\begin{tabular}{|ccccc|}

\tableline 
    Date    & $\Delta$t & Frequency & Flux Density &   Instrument   \\
            &  (days)   &   (GHz)   &   ($\mu$Jy)  &                \\
\tableline 
2005 Jan 29 &   671.90  &   4.86   & $324 \pm 28$ &   VLA-B$^1$    \\
2005 Mar 31	&   732.74	&   4.86	  & $196 \pm 31$ &	 VLA-A$^2$    \\
2005 Apr 07	&   739.73	&   4.86	  & $270 \pm 29$ &	 VLA-C$^1$    \\
2005 May 14	&   777.06	&   4.85	  & $276 \pm 38$ &	 WSRT$^5$     \\
2005 Jun 08	&   801.53	&   4.86	  & $279 \pm 27$ &	 VLA-A$^2$    \\
2005 Nov 27	&   974.51	&   4.80	  & $197 \pm 40$ &	 WSRT$^6$     \\
2006 Mar 22	&   1088.7 	&   4.86	  & $169 \pm 31$ &	 VLA-A$^3$    \\
2006 Apr 30	&   1128.0 	&   4.80	  & $149 \pm 34$ &	 WSRT$^7$     \\
2008 Feb 04	&   1773.3 	&   4.86	  & $70  \pm 17$ &	 VLA-A$^4$    \\
\tableline 

\end{tabular}
\end{center}
\caption{VLA and WSRT observations of the GRB 030329 radio afterglow at 5 GHz.  The data were taken under VLA project codes AF414$^1$, AK583$^2$, AS864$^3$, and TYP100$^4$, and WSRT sequence numbers 10502389$^5$, 10506025$^6$, and 10602115$^7$.  WSRT data appearing in this table correspond to data from \citet{vanderHorstEA05} and \citet{vanderHorstEA08} that were re-analysed to ensure uniformity of analysis across all data points.   \label{VLAobservations}}
\end{table}

\begin{figure}[h]
\centering
\includegraphics[scale=0.44]{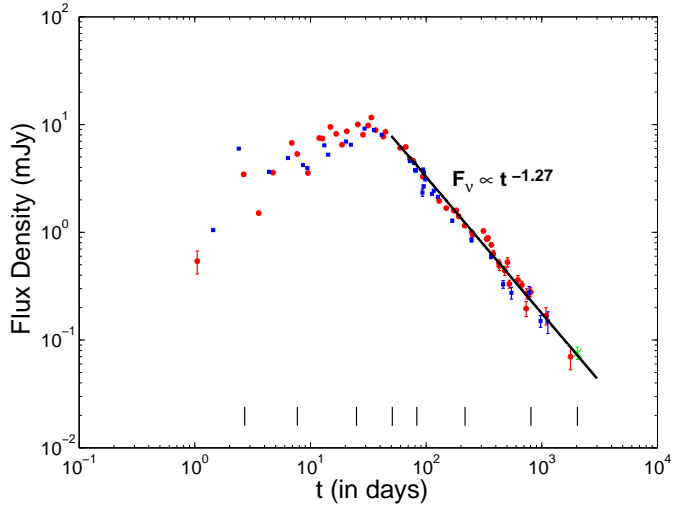}
\caption{The GRB 030329 5GHz light curve can be represented as a decaying power law with $\alpha = 1.27 \pm 0.03$ after day 59.  Red circles are data taken from the VLA archive, while blue squares are data taken from this work, \citet{vanderHorstEA05}, and \citet{vanderHorstEA08}.  The vertical lines at the bottom of the figure denote dates at which the afterglow size was measured with VLBI.  The green cross represents the VLBI data point presented in this work.  \label{lightCurve}}
\end{figure}

Using the VLA data included in this work in conjunction with previously-published VLA data \citep{frailEA05, bergerEA03a, pihlstromEA07} and WSRT data \citep{vanderHorstEA05, vanderHorstEA08}, we find that the decrease in flux density from day 59 to day 1773 (Fig. \ref{lightCurve}) is well described by a power law $F_\nu \propto t^{-\alpha}$ with the temporal index $\alpha = 1.27 \pm 0.03$. This is consistent with the value of $\alpha = 1.23 \pm 0.03$ obtained by \citet{pihlstromEA07} for the time range of 59 to 806 days after the burst.

\section{Results}


\subsection{Flux Density Evolution}

Our archival VLA and WSRT observations show that the evolution of the afterglow flux density follows a single power law drop-off of the form $F_\nu \propto t^{-\alpha}$ with a temporal index $\alpha = 1.27\pm0.03$ (see Fig.\ \ref{lightCurve}) from day 59 through day 1773.  This is consistent with previous work by \citet{pihlstromEA07}, who found that the drop-off was $F_\nu \propto t^{-1.23\pm0.03}$ between days 59 and 621.  Our VLBI measurement on day 2032 is also consistent with the overall power law, with an observed flux density of $76\pm10\ \mu$Jy closely matching the predicted flux of $73\ \mu$Jy.  VLBI instruments, lacking short baselines, are much less sensitive to extended emission than smaller interferometers such as the VLA and the WSRT.  In order for measurements of the flux density made with the VLA and the WSRT to agree with measurements taken from VLBI observations, there cannot be significant extended emission. 

The electrons that are injected into a gamma-ray burst shock are generally assumed to follow a power law distribution in energy $N(\gamma_e) \propto \gamma_e^{-p}$.  The temporal index can be used to determine the value of the electron power law index $p$ \citep{vanderHorstEA08}.  In the non-relativistic regime, the temporal index and the electron spectral index, $p$, are related by $\alpha = 3(5p - 7)/10$ in the homogenous case and by $\alpha = (7p - 5)/6$ in the wind case.  Our value of $\alpha = 1.27\pm0.03$ yields $p = 2.25\pm0.02$ for a uniform medium and $p = 1.80\pm0.03$ for a wind.  If the circumburst density is assumed to follow a power law in radius $\rho_{\rm ext} \propto r^{-k}$, then the slope power law $k$ can be calculated directly using

\begin{equation}
k = \frac{5\alpha - 15\beta + 3}{\alpha - 4\beta + 2}, \label{kEquation}
\end{equation}

\noindent where $\beta$ is the spectral index \citep{vanderHorstEA08}.  Using the value of $\beta = 0.54 \pm 0.02$ from \citet{vanderHorstEA08}, one finds $k = 1.1\pm0.2$.  It is important to note, however, that the value that is obtained for $p$ (and, therefore, $k$) using this method is extremely sensitive to the time range that is used in the calculation.  The transition of the jet to non-relativistic expansion leads to a steeper decay phase than in the Newtonian regime, and will lead to a higher value of $p$.  It is contended by \citet{vanderHorstEA08} that the jet becomes non-relativistic at $t \simeq 100\ \text{days}$, and so any data before this time is excluded.  With this choice of cutoff time, they obtain a value of $p = 2.12\pm0.02$ and $k = 0.33_{-0.41}^{+0.34}$.  They point to the fact that they obtain larger values of $p$ if they include earlier data to support their choice of the cutoff time.  Setting the cutoff to $t = 312\ \text{days}$, however, yields an even larger value of the electron power law index of $p = 2.43 \pm 0.07$ ($k = 1.9\pm0.19$), even though the burst is almost certainly entirely non-relativistic by this time.  Setting the cutoff time to $t = 426 \text{days}$, on the other hand, yields $p = 2.14 \pm 0.07$ ($k = 0.46\pm0.44$.  It is also noted by \citet{vanderHorstEA08} that the value of $\chi^2$ decreases significantly if data taken prior to $t = 100$ days are excluded, which they use as additional support of their choice of cutoff time.  Choosing a cutoff time of $t = 312$ days, however, will yield an even lower value of $\chi^2$ than a cutoff time of $t = 100$ days.  We find that the value of $k = 1.27\pm0.03$ is consistent with \citet{vanderHorstEA08} given the large uncertainty in $k$ that is obtained by calculating it directly from the temporal and spectral slopes.  It should also be noted however that \citet{vanderHorstEA08} used multi-frequency data to obtain their value of $p$, meaning that they had more independent data points with which to calculate it. 


\subsection{Size and Expansion Rate}

\begin{figure}[t]
\centering
\includegraphics[scale=0.44]{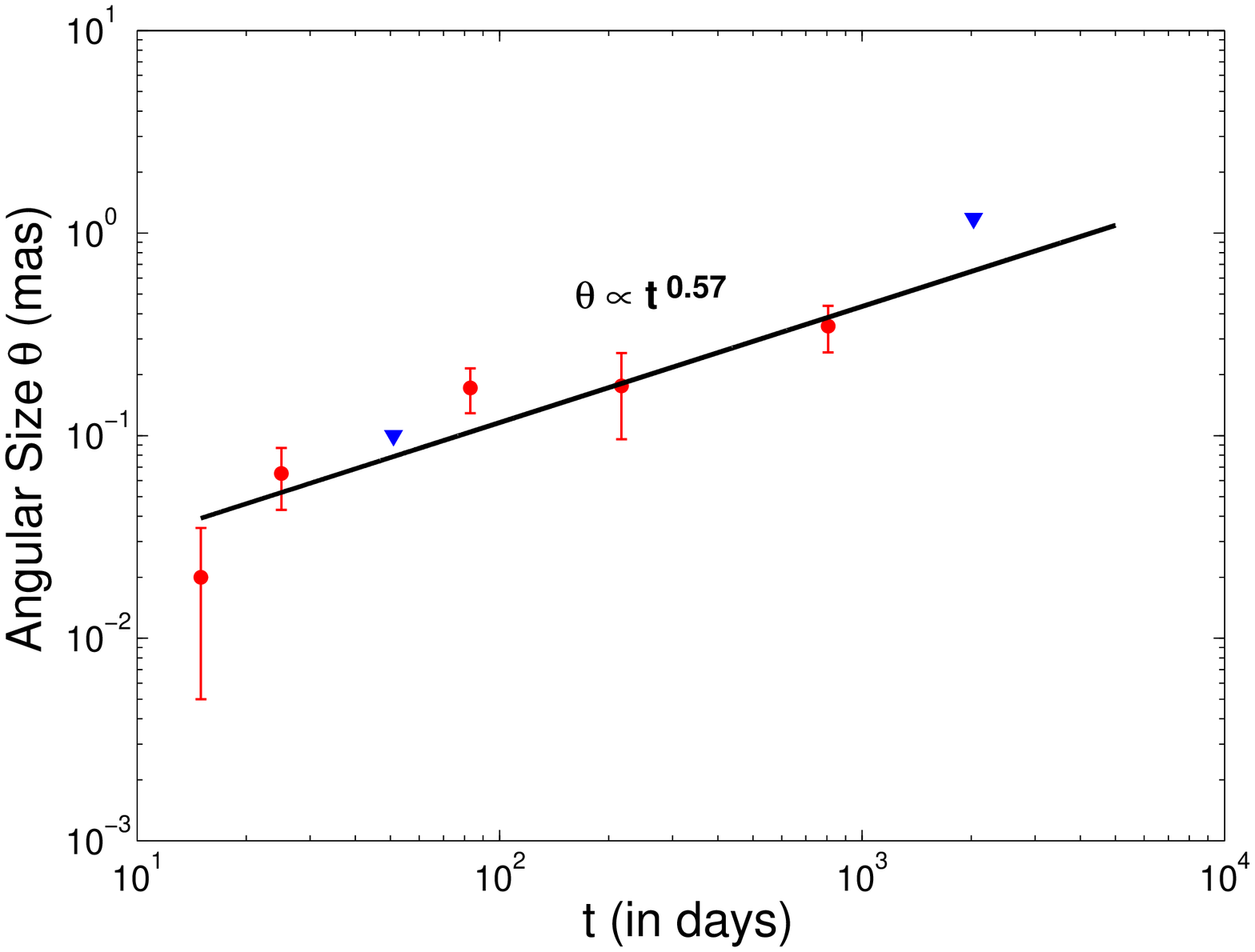}
\caption{The apparent expansion of GRB 030329 derived from measurements and limits
on the angular size as a function of time. The $1\sigma$ upper limit at day 51 is
from \citet{taylorEA04}, as are the measurements on days 25 and 83. The
measurement on day 217 is from \citet{taylorEA05}.  The measurement on day 806 comes from
\citet{pihlstromEA07}.  The measurement on day 15 comes from a model-dependent estimate based on the quenching of the scintillation \citep{bergerEA03a}.  Finally, the $1\sigma$ upper limit on day 2032 comes from this work.   \label{sourceSize}}
\end{figure}

The entire history of expansion for GRB 030329 is shown in Fig. \ref{sourceSize}. The first measurement at 15 days comes from a model-dependent estimate of the quenching of the scintillation \citep{bergerEA03a}. The uncertainties on this size estimate are large due to the dependence of the measurement on estimated properties of the interstellar medium along our line-of-sight.  A $2\sigma$ upper limit on the afterglow image size of 1.31 mas, or $1.2\times10^{19}$ cm, was obtained for day 2032.  

The apparent expansion speed is defined as $\beta_{\rm app}c$.  The decline in $\langle\beta_{\rm app}\rangle$ with time is shown in Fig. \ref{beta}, where $\langle\beta_{\rm app}\rangle$ is found using

\begin{equation}
\langle \beta_{\rm app} \rangle = \frac{(1+z)R_\perp}{ct} =\frac{\theta_R d_M}{ct},
\end{equation}

\noindent where $\theta_R = R_\perp/d_A = (1 + z)R_\perp/d_M$ and $R_\perp$ are the angular and physical radius of the image, respectively, $d_M$ is the proper distance to the source, $z$ is the source's cosmological redshift, $t$ is the time of observation, and $c$ is the speed of light.  Previous work by \citet{pihlstromEA07} and \citet{taylorEA04,taylorEA05} shows a continuous decrease in the apparent expansion rate of the image up to day 806, likely due to a deceleration of the afterglow shock due to interaction with the circumburst medium.  The mean afterglow apparent expansion rate up to day 806 was $\langle \beta_{\rm app} \rangle = 0.9 \pm 0.2$.  Our upper limit of 1.31 mas at day 2032 corresponds to a mean apparent expansion rate of $\langle \beta_{\rm app} \rangle < 1.3$.  It should be noted that the intrinsic image surface brightness profile of the afterglow is expected to have a small effect on the value obtained for the diameter of the afterglow $2R_{\perp}$.  The image may be limb-brightened \citep{granotEA99, granotLoeb01}, which could suggest a possible correction factor to the values used in this paper of $\sim1.4$.  For an in-depth discussion of the effects of surface brightness profiles on the afterglow diameter, see \citet{pihlstromEA07}. 

\subsection{Proper Motion}
Solving for proper motion using all the VLBI observations to date, we derive $\mu_{\rm r.a.} = 0.0044\pm0.054$ mas yr$^{-1}$ and $\mu_{\rm dec.} = -0.0031\pm0.033$ mas yr$^{-1}$, or an angular displacement over 2032 days of $0.030\pm0.35$ mas (Fig. \ref{skyPosFigure}). These observations are consistent with those reported by \citet{taylorEA04, taylorEA05} and \citet{pihlstromEA07}, and impose a stronger limit on the proper motion. The implied $2\sigma$ limit on the proper motion in the plane of the sky is $0.067$ mas yr$^{-1}$ (corresponding to $1.1$ pc) in $2032(1 + z)^{-1}\approx1739$ days, or $0.73c$ ($1\sigma$).

In the relativistic fireball model, a shift in the flux centroid is expected owing to the spreading of the jet ejecta \citep{sari99}. For a jet viewed off the main axis, the shift can be substantial \citep{granotLoeb03}. However, since gamma rays were detected from GRB 030329 it is likely that we are viewing the jet largely on-axis.  

The angular shift of the flux centroid is approximately $q\theta_0R/D$ \citep{sari99}, where $q$ is a measurement of the angular separation between the observer and the center of the jet in units of the half-opening angle of the jet, $\theta_0$ is the initial half-opening angle of the jet, $R$ is the radius of the emitting region, and $D$ is the angular distance to the observer.  The maximum angular shift is found by setting $q = 1$, in which case the observer is at the edge of the jet.  For a typical value of the initial jet half-opening angle of $\theta_0 = 0.2$ mas, our upper limit on the emitting region size of $R < 1.2\times10^{19}$ cm, and a luminosity distance of $D = 587$ Mpc, the maximum angular shift in the flux centroid is $0.16$ mas.  Although this estimate is larger than our upper limit of $0.067$ mas, it can easily be made consistent by either decreasing the value of $\theta_0$ or of $q$.  Indeed, the detection of a gamma-ray afterglow makes it likely that we are positioned near the center of the jet and $q \ll 1$.

Proper motion in the cannonball model originates from the superluminal motion of plasmoids ejected during a supernova explosion with $\Gamma_0 \sim 1000$ \citep{dadoEA03a}.  \citet{darDeRujula03} predicted a displacement of 2 mas by day 80 assuming plasmoids propagating in a constant density medium. This estimate was revised downward to 0.55 mas by incorporating plasmoid interactions with density inhomogeneities at a distance of $\sim100$ pc within a wind-blown medium \citep{dadoEA04}. Neither variant of this model is consistent with our proper-motion limits.

\section{Discussion}


\subsection{Density Profiles}

Fig.\ \ref{models} demonstrates how the measured evolution of the image size for GRB 030329 compares with the predictions of a suite of theoretical models developed by \citet{granotEA05}. The data constrain the external density profile at radii $R \gtrsim 10^{18}$ cm.  Our observation at day 2032 suffered from a very low signal to noise, resulting in a large uncertainty.  Additionally, the source is unresolved, meaning that we can only report an upper limit of $1.31$ mas ($2\sigma$). 

\begin{figure}[h]
\centering
\includegraphics[scale=0.44]{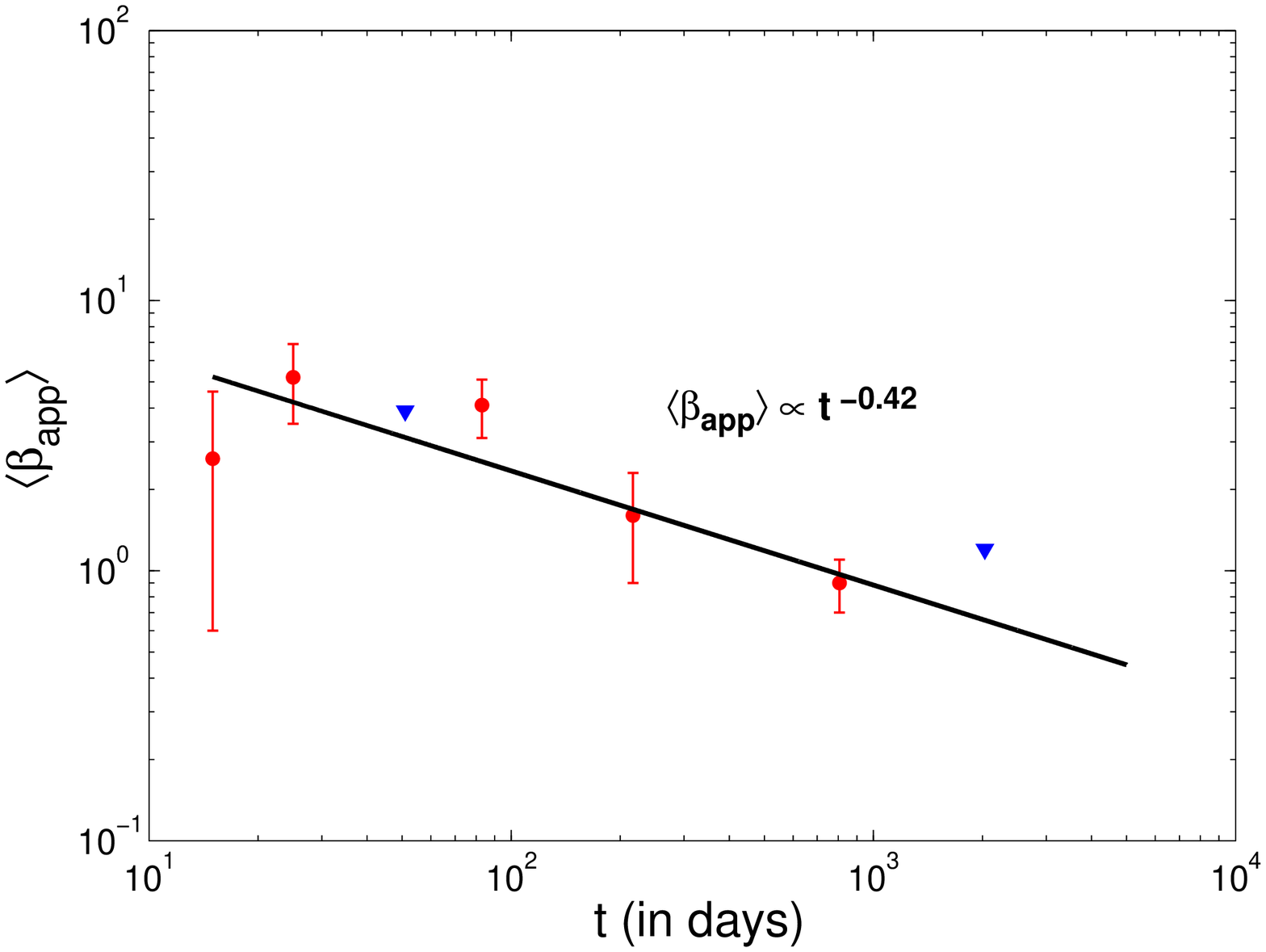}
\caption{Evolution of the average apparent expansion velocity derived from direct size
measurements, and assuming a Gaussian intrinsic surface brightness profile. The $1\sigma$ upper limit at day 51 is from \citealt{taylorEA04}, as are the measurements on days 25 and 83. The
measurement on day 217 is from \citealt{taylorEA05}.  The measurement on day 806 comes from
\citealt{pihlstromEA07}.  The measurement on day 15 comes from a model-dependent estimate based on the quenching of the scintillation \citep{bergerEA03a}.  Finally, the $1\sigma$ upper limit on day 2032 comes from this work.   \label{beta}}
\end{figure}

Looking at Fig. \ref{models}, it can be seen that the upper limit that we obtain for the image size on day 2032 cannot be used by itself to distinguish between a uniform and a windlike medium.  Another, tighter constraint on the angular size can be inferred from the apparent afterglow expansion rate $\langle \beta_{\rm app} \rangle$ (Fig. \ref{beta}) on physical grounds.  The afterglow expansion rate at day 806 is  $\langle \beta_{\rm app} \rangle = 0.9\pm0.2$.  Although it is plausible that $\langle \beta_{\rm app} \rangle$ might have remained constant between days 806 and 2032, there is certainly no plausible physical mechanism that would cause it to increase.  The maximum afterglow size at day 2032 can therefore be calculated by assuming a constant expansion rate of $\langle \beta_{app} \rangle = 0.9$ over this period.  The result is a maximum afterglow size of $0.90$ mas ($8.0\times10^{18}$ cm), significantly lower than the measured upper limit of 1.31 mas ($1.2\times10^{19}$ cm) and at the low end of the range of values that are compatible with the wind models. 

Because we only have an upper limit on $\langle \beta_{\rm app} \rangle$ at day 2032, we cannot immediately rule out a large deceleration due to the transition of the jet into a more dense environment such as the ISM that lies beyond a star's wind termination shock.  A non-relativistic jet that encounters an enhancement in the density is expected to produce a rebrightening in the light curve \citep{meslerEA12, ramirez-RuizEA01, ramirez-RuizEA05, daiLu02}.  The jet transitions to the non-relativistic regime when $\langle \beta_{\rm app} \rangle \simeq 2$ \citep{pihlstromEA07}, or after approximately 1 year.  The lack of any rebrightening in the 5 GHz light curve argues against an increasing density.  The observations of the afterglow size at days 217 and 806 are consistent with a uniform medium.  With limits on the possible afterglow expansion rate making a windlike medium unlikely and a lack of any rebrightening in the light curve making a transition to either a denser uniform medium or a $k>0$ density profile unlikely as well, it is most probable that the jet is propagating through a continuous, uniform medium.  In a follow-up paper we plan to use state-of-the-art relativistic hydrodynamic simulations \citep{deColleEA12,deColleEA11}, which would provide a significantly more realistic modeling of the GRB jet dynamics.


\subsection{Rebrightening}

Radio rebrightening of GRB 030329 has been predicted by \citet{granotLoeb03} as the counter jet becomes non-relativistic and its emission is no longer strongly beamed away from us. At 15 GHz, \citet{liSong04} predicted that the flux density of the afterglow would be 0.6 mJy 1.7 yr after the burst.  The actual 4.86 GHz flux density was $\sim0.33$ mJy after 1.7 yr (see Fig.\ \ref{lightCurve}).  Using the relationship $F_\nu \propto \nu^{-0.6}$, which is valid between these two frequencies, one would expect to find a corresponding flux density at 15 GHz of $\sim0.17$ mJy, a factor of $\sim3.3$ lower than the prediction.  A more recent model by \citet{vanEertenEA10} predicts that the contribution to the flux by the counter-jet should be significantly smaller and should occur much later at $\sim600$ days.   We find that the late time 4.9 GHz light curve, from 59 days to 2032 days, is consistent with a single power law decay of $F_\nu \propto t^{-1.27 \pm 0.03}$ (see Fig. \ref{lightCurve}), which implies that, up to $\simeq$ 2032 days after the burst, there is no significant contribution to the observed flux density from the counter jet.  Consequently, we find no clear evidence for a rebrightening up to 5.5 years after the burst.  We note that detailed numerical simulations show that the rebrightening due to the counter-jet becomes much less prominent for a wind-like density profile (see Fig. 13 of \citealt{deColleEA12}).  In such a case it might still be consistent with the observed single power-law flux decay in the late time radio afterglow light curve of GRB 030329. 


\section{Conclusions}
We present measurements of the 5 GHz flux density and image size of the GRB 030329 radio afterglow taken over a period of 5.5 years.  These observations clearly demonstrate that the expansion rate has decreased over time, with a transition to the non-relativistic regime at $\sim$ 1 yr.  After approximately day 59, the afterglow flux density follows a power law of $F_\nu \propto t^{-\alpha}$, with temporal index $\alpha = 1.27\pm 0.03$, which agrees with the value of $\alpha = 1.23\pm 0.03$ obtained by \citep{pihlstromEA07}.  Using the method of \citet{vanderHorstEA08}, the electron power law index $p$ is found to be $p = 2.24\pm0.02$ for a uniform medium, which does not agree with their value of $p = 2.12\pm0.02$.  The value determined for the electron power law index is found to be highly sensitive to the time range used to calculate it, making the temporal slope only capable of providing a quick estimate of the electron power law index.


A rebrightening of the source was expected as the counter jet became non-relativistic, however, no rebrightening was detected up to 5.5 years after the burst.  Numerical simulations suggest that any rebrightening in a windlike medium would be more difficult to detect as compared to a uniform medium.  However, a windlike medium is not favored by the measured evolution of the afterglow image size.  A possibility also exists for asymmetry between the jet and the counter jet.  If the counter jet had a lower initial energy or encountered a less dense external medium, then it would produce a smaller rebrightening than expected. 

An upper limit of $0.067$ mas yr$^{-1}$ is found for the proper motion.  Consequently, the proper motion of the flux centroid is constrained to be smaller than the diameter of the image, which is consistent with the fireball model but not with the cannonball model. 

\begin{figure}[t]
\centering
\includegraphics[scale=0.65]{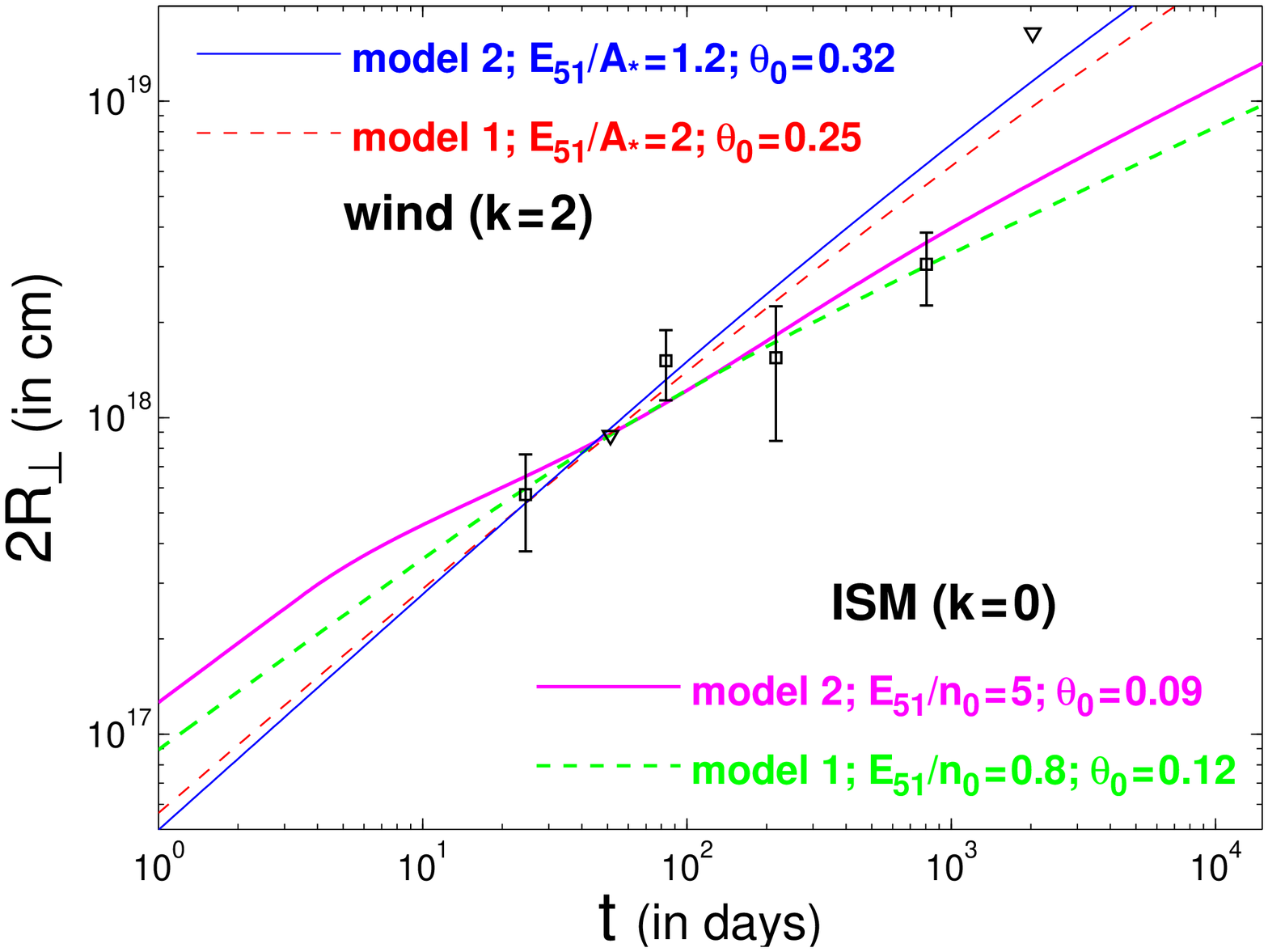}
\caption{Tentative fits of theoretical models for the evolution of the source size (from \citealt{granotEA05}) to the observed image size (of diameter $2R_\perp$) of the radio afterglow of GRB 030329 up to 83 days.  The $1\sigma$ upper limit at day 51 is from \citealt{taylorEA04}, as are the measurements on days 25 and 83. The measurement on day 217 is from \citealt{taylorEA05} and the measurement at 806 days is from \citealt{pihlstromEA07}).  The $2\sigma$ upper limit at day 2032 comes from this work.  All direct measurements are plotted with 1$\sigma$ error bars. In model 1 there is relativistic lateral spreading of the GRB jet in its local rest frame, while in model 2 there is no significant lateral expansion until the jet becomes non-relativistic. The external density is taken to be a power law with the distance r from the source, $\rho_{ext} = Ar^{-k}$, where k = 0 for a uniform external density while k = 2 is expected for a stellar wind environment. \label{models}}
\end{figure}

\begin{figure}[b]
\centering
\includegraphics[scale=0.65]{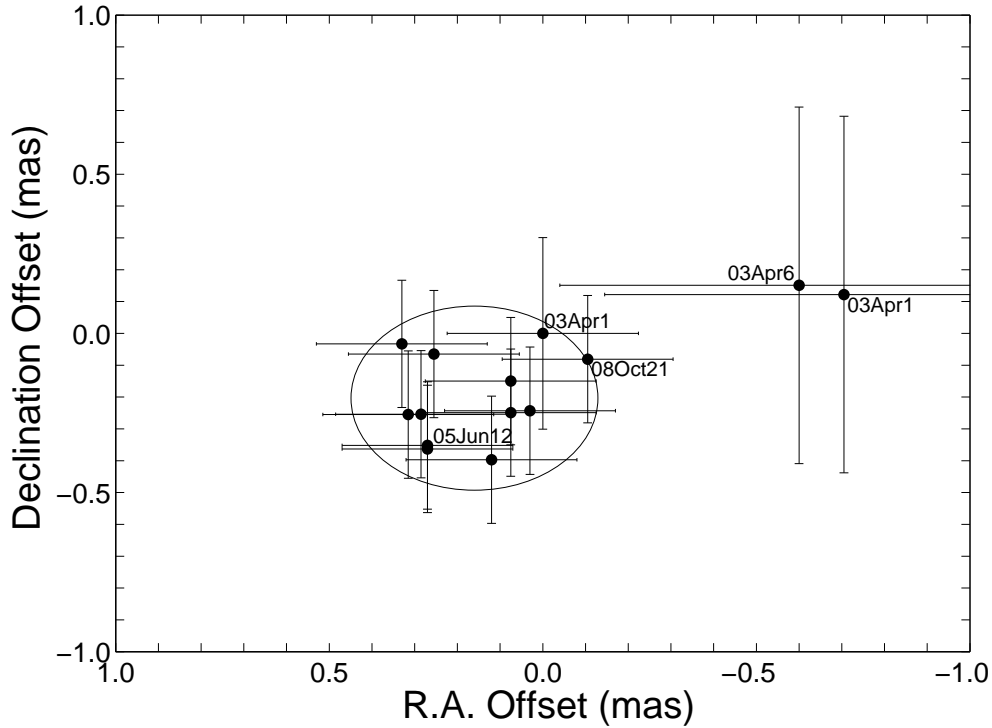}
\caption{The positions derived from the observations in eight epochs relative to the first determination on April 1st at 8.4 GHz. Observations at multiple frequencies at a given epoch have been plotted separately since they are independent measurements. A circle with a radius of 0.29 mas ($2\sigma$) is shown to encompass all measurements except for the one taken on 08 Oct. 21 (day 2032) which lies just outside the circle's edge, and those taken within
the first eight days at 5 GHz, which suffer from systematic errors \citep{taylorEA04}. Taken together these observations provide a constraint on the proper motion of $<0.067$ mas yr$^{-1}$ over 2032 days.  \label{skyPosFigure}}
\end{figure}

The upper limit that we obtained for the source size at day 2032 (Fig.\ \ref{models}) is, by itself, insufficient to conclusively determine the nature of the circumburst medium through which the jet is propagating.  We argue that, on physical grounds, the mean expansion rate $\langle\beta_{\rm app}\rangle$ cannot increase between two epochs.  The value of $\langle\beta_{\rm app}\rangle$ at day 806 was $0.9\pm0.2$, which corresponds to a maximum angular size at day 2032 of 0.90 mas, which is not consistent with a windlike medium.  We therefore argue that the expansion rate of the burst favors a uniform medium.  Moreover, a non-steady wind might produce an intermediate density profile with $k\sim 1$ where $\rho_{\rm ext} = Ar^{-k}$. This might potentially enable us to reconcile between the model and the afterglow image size evolution that favor a more uniform external medium, and the radio light curve that favors a more wind-like external density because of the lack of a bump in the light curve due to the counter-jet. This will be tested in a follow-up paper.  Future work includes a comparison of our observations to detailed numerical simulations of the jet dynamics in different external density profiles \citep{deColleEA12, deColleEA11}.  Additionally, direct measurements of the afterglow expansion rate will be conducted on any future gamma-ray burst that is sufficiently bright to be observed with VLBI. 

\begin{acknowledgements}The National Radio Astronomy Observatory is a facility of the National Science Foundation operated under cooperative agreement by Associated Universities, Inc.  The European VLBI Network is a joint facility of European, Chinese, South African and other radio astronomy institutes funded by their national research councils. The Arecibo Observatory is operated by SRI International under a cooperative agreement with the National Science Foundation (AST-1100968), and in alliance with Ana G.\ M\'endez-Universidad Metropolitana, and the Universities Space Research Association. The Effelsberg 100-m telescope is operated by Max-Planck-Institut f\"ur Radioastronomie in Bonn (MPIfR) on behalf of the Max-Planck-Geselschaft. The Green Bank Telescope is operated by the National Radio Astronomy Observatory, a facility of the National Science Foundation operated under cooperative agreement by Associated Universities, Inc.
\end{acknowledgements}


\nocite{bergerEA03a}			\nocite{bergerEA03b}
\nocite{bergerEA03c}
\nocite{bloomEA03}				\nocite{chevalierLi00}
\nocite{daiLu02}	  			\nocite{deColleEA12}
\nocite{deColleEA11}			\nocite{frailEA01}
\nocite{frailEA05}				\nocite{FryerEA99}
\nocite{genetEA07}				\nocite{gorosabelEA06}
\nocite{granot05}		  		\nocite{granot07}
\nocite{granotEA99}				\nocite{granotEA03}
\nocite{granotEA05}				\nocite{granotLoeb01}
\nocite{granotLoeb03}			\nocite{granotsari02}
\nocite{greinerEA03}			\nocite{liSong04}
\nocite{meslerEA12}				\nocite{reesMeszaros94}
\nocite{meszarosRees97}		\nocite{mirabalHalpern06}
\nocite{orenEA04}					\nocite{paczynski93}
\nocite{pihlstromEA07}
\nocite{ramirez-RuizEA01}	\nocite{ramirez-RuizEA05}
\nocite{taylorEA04}				\nocite{taylorEA05}
\nocite{vanderHorstEA08}  \nocite{petersonPrice03}
\nocite{piranEA04}				\nocite{rhoads99}
\nocite{sariEA98}					\nocite{shethEA03}
\nocite{uhmBeloborodov07}	\nocite{vanderspekEA03}
\nocite{waxman97}					\nocite{meszarosRees93}  
\nocite{katz94}						\nocite{paczynski98}
\nocite{ramirez-RuizEA05} \nocite{vanEertenEA10}
\nocite{dadoEA03a}				\nocite{sari99}
\nocite{darDeRujula03}			\nocite{vanderHorstEA05}

\bibliographystyle{apj}
\bibliography{masterBibliography}

\end{document}